\documentclass[aps,twocolumn,prl,showpacs,preprintnumbers,amsmath,amssymb]{revtex4-1}

\usepackage{graphicx}
\usepackage{dcolumn}
\usepackage{bm}
\usepackage{txfonts}
\usepackage{color}

\begin{document}
\title{Reply to the Comment by S. Wirth {\it et al.} on ``Tuning low-energy scales in YbRh$_2$Si$_2$ by non-isoelectronic substitution and pressure''}

\author{M.-H.~Schubert$^{1}$}
\author{Y.~Tokiwa$^{2}$}
\author{S.-H.~H\"{u}bner$^{1}$}
\author{M. Mchalwat$^{1}$}
\author{E. Blumenr\"{o}ther$^{1}$}
\author{H.S.~Jeevan$^{1}$}
\author{P. Gegenwart$^{2}$}
\affiliation{$^{1}$I. Physik. Institut, Georg-August-Universit\"{a}t G\"{o}ttingen, D-37077
G\"{o}ttingen\\$^{2}${Experimental Physics VI, Center for Electronic Correlations and Magnetism, University of Augsburg, 86159 Augsburg, Germany}}

\date{\today}


\begin{abstract}
Previously, we reported that the doping and pressure dependence of the $T^\ast(B)$ crossover in YbRh$_2$Si$_2$ is incompatible with its interpretation as signature of a Kondo breakdown [M.-H. Schubert {\it et al.}, Phys. Rev. Research {\bf 1}, 032004(R) (2019)]. The comment by S. Wirth {\it et al.} [arXiv:1910.04108] refers to Hall measurements on undoped YbRh$_2$Si$_2$ and criticizes our study as incomplete and inconclusive. We thoroughly inspect these data and rebut the arguments of the comment. 
\end{abstract}

\maketitle

The heavy-fermion metal YbRh$_2$Si$_2$ has been intensively studied in the past 20 years because of its pronounced non-Fermi liquid effects near a quantum critical point (QCP) related to the suppression of very weak antiferromagnetic (AF) order ($T_{\rm N}=70$~mK) by a small critical magnetic field $B_{\rm c}$~\cite{Trovarelli,Gegenwart02}. In addition to the AF phase boundary, a crossover line $T^\ast(B)$, terminating for zero temperature at $B_{\rm c}$ was observed~\cite{Gegenwart05,Gegenwart07} and from its signature in the Hall effect a Kondo breakdown has been proposed~\cite{Paschen04}. A critical inspection of this interpretation is given below. In our previous work, we reported the combined influence of non-isoelectronic substitution and hydrostatic pressure on the low-temperature magnetoresistance, specific heat and magnetic susceptibility at fields applied within the magnetic easy plane perpendicular to the tetragonal c-axis~\cite{Schubert}. We found that $T^\ast(B)$ is a signature of field-driven moment polarization, insensitive to the balance of Kondo to RKKY interaction and with {\it finite} full width at half maximum (FWHM) in the $T \rightarrow 0$ limit, questioning the Kondo breakdown interpretation.

In their comment, S. Wirth {\it et al.}~\cite{Wirth} argue as follows: Previous Hall measurements on undoped YbRh$_2$Si$_2$ would have unambiguously proven a Kondo breakdown~\cite{Paschen04,Friedemann10} and our study would be inconclusive, because we did not measure Hall effect and did not properly take into account the effect of disorder, introduced by the doping. Below, we reply to this criticism by first focusing on the previous Hall data and subsequently on the effect of disorder in our doping study.\\

{\it Previous Hall measurements on undoped YbRh$_2$Si$_2$.}
\\

Fig.~1 of the comment by Wirth {\it et al.} \cite{Wirth} displays the temperature dependence of the FWHM of the Hall-crossover data from Ref.~\cite{Friedemann10}. A red line in this plot displays a linear temperature dependence. This line implies for the $T=0$ extrapolation a sudden jump of the Hall coefficient, that was associated with a Fermi surface change due to the Kondo breakdown. However, the data in this plot and their error bars are not as conclusive as commonly believed and stated in the comment. We demonstrate this by a few examples at temperatures below 0.1~K, which are most relevant for the conclusion on a zero crossover width at $T=0$.

\begin{figure}[t]
\includegraphics[width=0.9\linewidth,keepaspectratio]{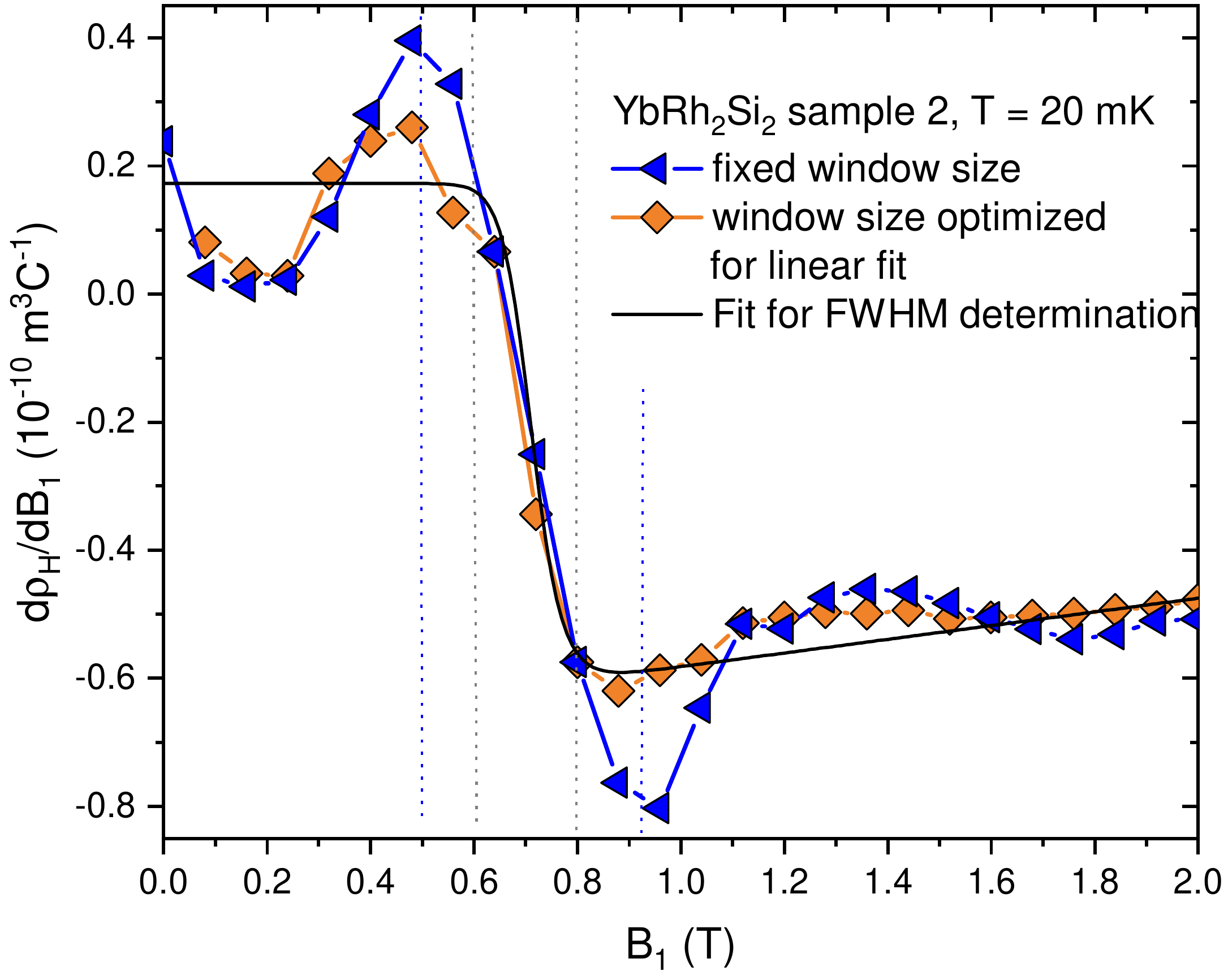}
\caption{Analysis of Hall data from ``single-field" measurements on sample 2 of~\cite{Friedemann10,Friedemann_PhD}. Blue triangles and orange diamonds result from calculations of the field-derivative utilizing a constant field window and field window of variable size, determined to optimize linearity in $B$~\cite{Friedemann_private}, as published in~\cite{Friedemann_PhD} and ~\cite{Friedemann10}, respectively. The fit for the determination of the FWHM~\cite{Friedemann10} is shown as black line. The grey and blue dotted lines display the boundaries of the crossover according to the fit and the data, respectively.}
\end{figure}

We first focus on single-field Hall measurements on sample 2 of~\cite{Friedemann10}. This is the sample with largest residual resistivity ratio of that study. From those measurements, the FWHM values represented by blue open circles in Fig. 1 of the comment~\cite{Wirth} were determined. At 20~mK, a FWHM value as small as 0.01~T has been indicated (since the field in the measurements was applied along the hard axis, this value was obtained by division of the determined FWHM by a factor 11, which takes into account the magnetic anisotropy~\cite{Friedemann10}). Fig.~1 focuses on the underlying data. It compares two sets of the numerically derived differential Hall coefficient $d\rho_{\rm H}(B_1)/dB_1$ data. The orange diamonds are from S. Friedemann {\it et al.}~\cite{Friedemann10} while the blue triangles are from the PhD thesis of Sven Friedemann~\cite{Friedemann_PhD}. We emphasize that the two sets of differential Hall coefficient curves are derived from identical data of  $\rho_{\rm H}(B_1)$. For the blue data set a fixed window for the differential derivative was used, while for the orange diamonds (from Fig.~1 of \cite{Friedemann10}) the numerical derivative was calculated with an algorithm that finds the window size that minimizes the standard deviation from linear fits~\cite{Friedemann_private}. In both cases, the description of the data by the empirical crossover-function (black line)~\cite{Friedemann10} is unsatisfactory, since the data show a couple of extrema, which are absent in this function. As a result of the advanced algorithm for the numerical derivative the ``oscillations" of the data around the fit by the crossover function are damped, but still it is evident, that the crossover function does not properly describes the data. It should be noted that at least some part of the oscillation is not just noise but an intrinsic feature visible {\it only} at temperatures below $T_{\rm N}$. Therefore, the fitted cross-over function is not suitable for temperatures below $T_{\rm N}$ and the data do not allow to conclude a proper crossover FWHM value. The fitted crossover appears between 0.6 and 0.8~T (cf. grey dotted lines). On the other hand, the blue dotted lines indicate the crossover visible in the actual data, which starts at 0.5~T and ends at 0.9~T. This yields an at least twice as large width, questioning the blue data points at lowest temperatures in Fig. 1 of the comment~\cite{Wirth}. However, as stated above, the fit function is not appropriate at low temperatures and thus it is difficult to conclude a sharpening of the Hall crossover below the AF ordering temperature.

\begin{figure}[t]
\includegraphics[width=0.9\linewidth,keepaspectratio]{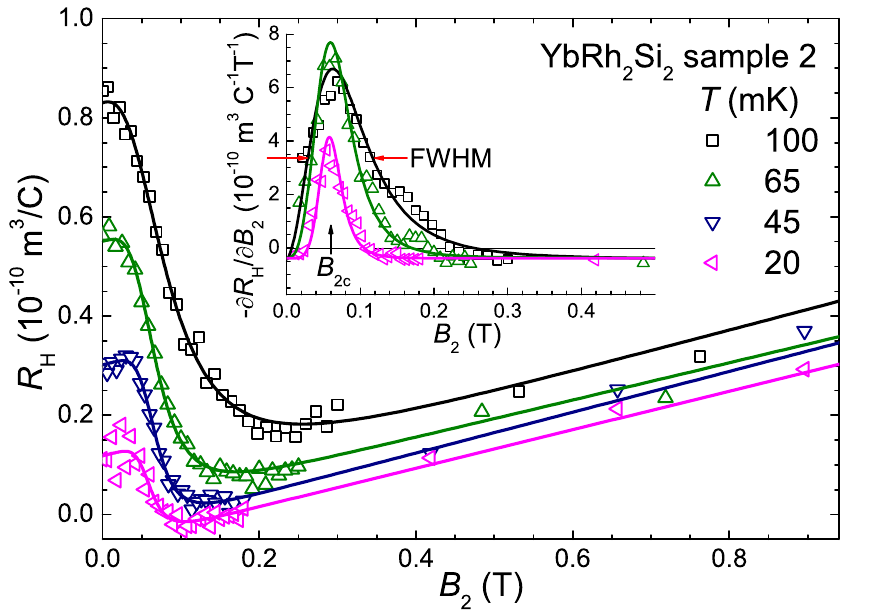}
\caption{Crossed-field Hall data from \cite{Friedemann10}, see text.}
\end{figure}

Next, we peruse the crossed-field Hall measurements, where $R_{\rm H}$ was determined from the initial slope of $d\rho_{\rm H}/dB_1$. Fig. 2 (adapted from ~\cite{Friedemann10}) shows $R_{\rm H}$ as function of the ``tuning" field $B_2$, parallel to the applied electrical current in the main panel. Within the scatter of the displayed $R_{\rm H}$ data (ignoring the fitted lines), the crossover at 20 mK does not appear 2.25 (3.25) times narrower than that at 45 (65) mK, required if the FWHM follows a proportionality with temperature. More likely, the $R_{\rm H}$ data indicate a saturation of the FWHM upon cooling to temperatures below 45~mK.
(The field-derivative of these data is shown in the inset with much weaker noise, probably resulting from sophisticated numerical data analysis. However, the differences in the crossover widths at lowest temperatures are not obvious from the raw $R_{\rm H}$ data.)


\begin{figure}[t]
\includegraphics[width=0.9\linewidth,keepaspectratio]{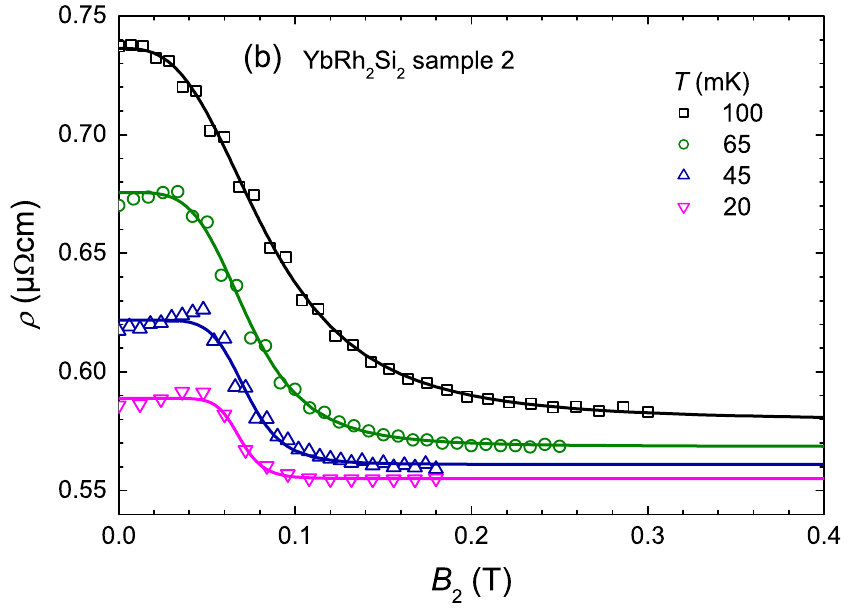}
\caption{Crossover in the magnetoresistance data from \cite{Friedemann10}, see text.}
\end{figure}

The FWHM vs. $T$ plot~\cite{Friedemann10} from Fig. 1 of the comment
by S. Wirth {\it et al.} also includes several points extracted from magnetoresistance measurements. However, as can be seen in Fig.~3, taken from the supplement of \cite{Friedemann10}, the magnetoresistance analysis at low temperatures is also problematic. There appears a maximum in $\rho(B)$ at temperatures below 70 mK, which interferes with the crossover. Since the fitting function improperly describes the data at the onset of the crossover, the true crossover width can well be larger than the value obtained by the fitting. One may argue that at elevated temperatures FWHM $\propto T$ without offset. However, at 100~mK the FWHM is almost 0.1~T, i.e., even larger than $B^\ast$ itself. Thus, there is a huge uncertainty extrapolating all the way down to $T=0$. In our work~\cite{Schubert}, we only included such data points to a FWHM {\it vs}. $T$ plot that were determined for the various doped samples above their respective $T_{\rm N}$ and for which the fit function describes the data without any such problems. The determined FWHM values are shown in Fig. 3(b) of our paper. A clear deviation from a $T$-linear behavior of the FWHM below 100~mK is visible, implying a finite crossover width at $T=0$. The comment emphasizes "the scattering rate has no way to create a jump"~\cite{Wirth}. Indeed, but such jump has not been unambiguously proven. Note, that within the theory of spin-flip scattering of critical quasiparticles by Wölfle, Abrahams and Schmalian, the crossover signatures at $B^\ast$ in YbRh$_2$Si$_2$ and their thermal broadening are accounted for by the Zeeman effect without invoking a Kondo breakdown~\cite{WA,WAS}. \\

{\it Difficulties of results on undoped YbRh$_2$Si$_2$ with the Kondo breakdown scenario}
\\

The comment argues, magnetoresistance measurements alone would not allow us to draw conclusions on the nature of $T^\ast$~\cite{Wirth}. Supposing a drastic change of the Fermi surface at $T=0$, however, all physical properties related to heavy quasiparticles should display respective signatures. In fact, the magnetoresistance change at the $B^\ast$ in YbRh$_2$Si$_2$ has been argued by the authors of the comment to be caused by a Kondo breakdown, previously~\cite{Smidman}.

As stated before, the crossover is still broad at 0.1~K and a truly $T$-linear behavior of the width, that would extrapolate to a jump in the zero-temperature limit, is not evident from the available data. Another problem with the Kondo-breakdown interpretation of the $T^\ast$ crossover is discussed in the following.

\begin{figure}[t]
\includegraphics[width=0.95\linewidth,keepaspectratio]{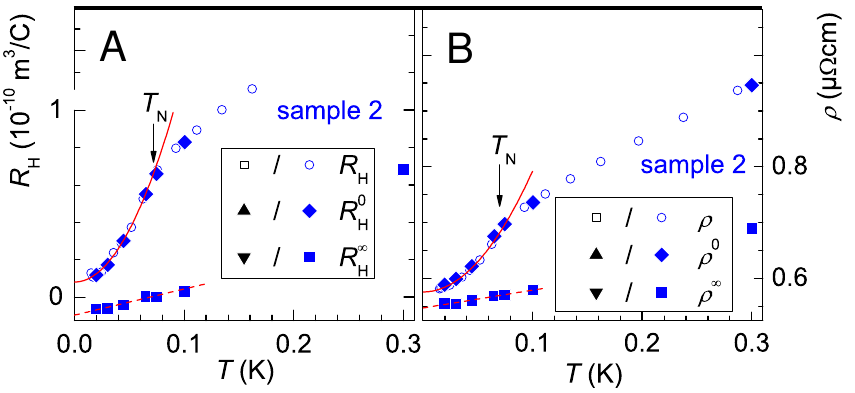}
\caption{Temperature dependence of Hall coefficient (A) and electrical resistance (B) for YbRh$_2$Si$_2$ (from \cite{Friedemann10}).}
\end{figure}

The Fermi surface of YbRh$_2$Si$_2$ was recently investigated by angular resolved photoemission spectroscopy (ARPES)~\cite{Kummer}. Down to 1~K, a large Fermi surface was detected at zero field, while the Kondo breakdown scenario expects a small Fermi surface at $B<B^\ast$. Furthermore, scanning tunneling spectroscopy (STS) measurements at 300 mK did not detect significant differences in the spectra for field smaller and larger than $B^\ast$, indicating a dominant signature of the large Fermi surface at zero field~\cite{Seiro}. It has been argued~\cite{Paschen16,Smidman}, that in view of the substantial crossover width down to these temperatures, physical properties being sensitive to the Fermi surface (such as ARPES, STS, Hall-effect, etc.) would see at ``elevated" temperatures significant spectral weight of the large Fermi surface at zero magnetic field, even though the zero-field state is claimed to be the one with small Fermi surface. Consequently, signatures implying differences between large and small Fermi surface, invisible down to 300 mK, should become visible upon cooling, as the crossover width decreases linearly with $T$ and spectral weight of the large Fermi surface thus is expected to disappear at zero field. However, the temperature dependence of the Hall constant at both sides of the $B^\ast$ crossover, displayed in Fig. 4(A)~\cite{Friedemann10}, indicates disparate behavior: The change of the Hall constant, $R_{\rm H}^0$-$R_{\rm H}^\infty$, decreases from 300 mK to 20 mK by a factor of 5. If the spectral weight of the large Fermi surface would freeze out at zero field upon cooling, one may have expected to see $R_{\rm H}^0$-$R_{\rm H}^\infty$ increasing upon cooling. In fact, the change of the Hall constant with temperature very much resembles that of the electrical resistance (cf. Fig. 4(B)). This indicates that both properties are mainly governed by scattering and not by a presumed change of charge carrier concentration~\cite{Paschen04}. Thus, also the combination of ARPES, STS and Hall effect questions a Kondo breakdown at $B^\ast$ in undoped YbRh$_2$Si$_2$.\\


The scenario of a locally critical QCP discusses that the transition from a small to a large Fermi surface is driven by enhancing the relative strength $J_{\rm K}/J_{\rm RKKY}$ of the Kondo compared to the RKKY coupling~\cite{Si}. Note, that magnetic field has not been explicitly included to an underlying Hamiltonian. For the $T^\ast$ anomaly in YRS it is assumed that field tilts the balance between these two scales in favor of the larger Fermi surface. However, if $T^\ast$ would depend on the competition between $T_{\rm K}$ and $T_{\rm RKKY}$ it should display a pressure dependence (as clearly observed for the AF phase boundary~\cite{Trovarelli}). Such pressure or chemical pressure dependence of $T^\ast$ is however almost absent~\cite{Friedemann,Tokiwa09}, despite huge opposite changes of $T_{\rm N}$ and the single-ion Kondo temperature $T_{\rm K}$. This indicates, that the $T^\ast$ crossover is almost insensitive to the balance of Kondo to RKKY interaction. It was noticed that the pressure insensitivity of $B^\ast$ at $T=0$ can be mapped to a global phase diagram, if pressure would only act in reducing the parameter $G$ (i.e., the strength of quantum fluctuations, induced e.g. by geometrical frustration), leaving $J_{\rm K}/J_{\rm RKKY}$ unchanged~\cite{Steglich_Si}. However, there is neither experimental indication (see above) nor theoretical justification for a pressure insensitivity of $J_{\rm K}/J_{\rm RKKY}$.\\

{\it Effect of disorder in doped YbRh$_2$Si$_2$}
\\

Finally, we reply to the criticism concerning a possible effect of disorder in our study of doped YbRh$_2$Si$_2$. We carefully examined our substituted single crystals by XRD and EDX and confirmed phase purity, homogeneity and actual compositions (cf. the supplement of~\cite{Schubert}). The residual resistivity of Fe- and Ni-substituted crystals is similar to that of respective Co- and Ir-substitutions, for example about $7~\mu\Omega$cm for $x\approx 0.07$~\cite{Schubert,Friedemann}. Note, that for Co- and Ir-substitutions $B^\ast$ has also been determined by very similar magnetoresistance crossovers, which have been interpreted as signatures of the Kondo breakdown being detached from the AF QCP. Furthermore, for the various Fe- and Ni-substitutions, the characteristic maximum of the electrical resistivity, which arises from the single-ion Kondo effect, as well as the size of $T_{\rm K}$, determined from the magnetic entropy, follow the previous trend found for Co-substitution~\cite{Friedemann}, indicating a similar chemical pressure effect~\cite{Schubert}. Moreover, application of hydrostatic pressure on Yb(Rh$_{0.93}$Fe$_{0.07}$)$_2$Si$_2$ leads to a similar stabilization of AF order (and very similar electrical resistivity signatures of the latter), as previously found in pressure experiments on undoped as well as Co-doped YbRh$_2$Si$_2$ (considering of course the respective differences in chemical pressure)~\cite{Friedemann_PhD}. The combination of these results provides confidence, that the disorder effects introduced by Fe- or Ni-substitution are under control and similar as for Co- and Ir-substitutions. Indeed, the effect of our substitution “cannot be captured by a change in the residual resistivity alone” as stated in the comment~\cite{Wirth}. This has been one of our main points~\cite{Schubert}: Partial Fe-substitution suppresses the $T^\ast$ crossover and the ferromagnetic correlations, while Ni-substitution acts oppositely. The partial moment polarization at $B^\ast$ is directly visible in magnetization, magnetic susceptibility and magnetic entropy~\cite{Tokiwa09,Gegenwart05,Tokiwa,Lausberg}, indicating the field-driven nature of this crossover.\\

{\it Conclusion}
\\

Several observations in YbRh$_2$Si$_2$, including the stronger than logarithmic quasiparticle mass divergence~\cite{Trovarelli,Gegenwart02,Custers}, Gr\"uneisen parameter divergences with fractional exponents~\cite{Kuechler,Tokiwa} and the linear temperature dependence of the electrical resistivity~\cite{Gegenwart02,Custers} are clearly incompatible with the predictions of the itinerant theory for an AF QCP, while they have been accounted for by a theory of critical quasiparticles~\cite{WAS}. This theory also quantitatively describes the Hall effect and magnetoresistance behavior of pure YbRh$_2$Si$_2$ by the Zeeman effect acting on robust critical quasiparticles~\cite{WA}. On the other hand, the theory of a locally critical QCP~\cite{Si} also accounts for the above non-Fermi liquid effects and treats the $T^\ast(B)$ crossovers, also for Yb(Rh$_{1-x}$T$_x$)$_2$Si$_2$ with T=Co or Ir, where they are clearly separated from the AF QCP, as signatures of the Kondo breakdown~\cite{Steglich_Si}. However, this interpretation seems incompatible with the experimental observations discussed above and in ~\cite{Schubert}. Clearly $T^\ast(B)$ reflects a partial magnetic polarization, indicated by the signature in the differential magnetic susceptibility. Field-driven crossovers were also found in other heavy fermion metals~\cite{Ce3Pd20Si6,Zhang,CeRhIn5,CePdAl2019}.
Note that our discussion does not exclude a Kondo breakdown in other situations, e.g. for pressurized CeRhIn$_5$~\cite{Shishido} or substituted CeCu$_{6-x}$Au$_x$~\cite{Schroeder}.

To summarize, a critical inspection of the $T^\ast(B)$ crossover signatures in undoped YbRh$_2$Si$_2$ reveals clear contradiction to the expectation of a Kondo breakdown QCP. This includes its pressure insensitivity and the saturation of the FWHM, being incompatible with a sharp Fermi surface jump at $T=0$. The crossover signatures are naturally related to a moment polarization and thus their theoretical description requires the consideration of the Zeeman effect.\\



{\it Acknowledgment}
\\

We thank S. Friedemann for providing original data and accompanying information, as well as M. Brando, C. Geibel, J. Schmalian and P. W\"{o}lfle for useful discussions.

\end{document}